\documentclass{aa}
\usepackage{graphicx}
\usepackage{txfonts}
\usepackage{epsfig}
\usepackage{times}
\def\ltsima{$\; \buildrel < \over \sim \;$}
\def\gtsima{$\; \buildrel > \over \sim \;$}
\def\simlt{\lower.5ex\hbox{\ltsima}}
\def\simgt{\lower.5ex\hbox{\gtsima}}

\begin{document}
   \title{X-ray absorption in Compton--thin AGN: the predictions of a model revisited}
   \author{Alessandra Lamastra, G. Cesare Perola and Giorgio Matt}
   \offprints{lamastra@fis.uniroma3.it}
   \institute{Dipartimento di Fisica ``E. Amaldi'', Universit\`a degli Studi Roma Tre,
via della Vasca Navale 84, I-00146 Roma, Italy}
   \date{Received ; Accepted }
   \abstract{The evidence of a decrease with increasing luminosity in the fraction $f_{abs}$
of absorbed and Compton-thin among X-ray-selected (2-10 keV) AGN is observationally well supported,
while that of an increase in $f_{abs}$ with redshift is fairly
controversial. In Lamastra, Perola \& Matt (2006), the
gravitational effect of the SMBH on the molecular interstellar gas, in the central region of
the host galaxy, was shown to predict an anti-correlation between $f_{abs}$
and the black-hole mass $M_{BH}$.}
{The most recent findings on the distribution of the Eddington ratio $\lambda$=$L_b/L_E$ as a function
of $M_{BH}$ and $z$ are used to convert that relationship into one between $f_{abs}$
and both  bolometric ($L_b$) and X-ray ($L_X$) luminosities at various values of $z$.}{The findings for $\lambda(M_{BH},z)$ are
properly treated to ensure completeness in the prediction of $f_{abs}$ above
a certain luminosity, at values of $z$=0.1, 0.35, 0.7, and $\geq$1. To verify the consequence
of these findings alone, we first adopted a distribution of gas surface density $\Sigma$, observed
in a sample of local spiral galaxies, irrespective of the galaxy morphological type and $z$.}{Assuming the Eddington limit, $\lambda$=1, in the $\lambda(M_{BH},z)$ distribution  as a ``natural'' cut-off, the predictions are consistent
with the existence of an anti-correlation between $f_{abs}$ and $L_X$, but they fail
to reproduce an increase in $f_{abs}$ with $z$. Because the early type galaxies
are on average much poorer in molecular gas than late type ones, a quantitative 
agreement with the local value of $f_{abs}$ requires the existence of a correlation
between $\Sigma$ and the central activity. An increase in typical values of $\Sigma$
with $z$, correlated with the activity, might explain an increase in $f_{abs}$ with $z$. However, $f_{abs}$ could 
hardly exceed about 0.3 at the highest luminosities.}{}


\authorrunning{A. Lamastra, G.C. Perola and G. Matt}

\titlerunning{Compton--thin type 2 AGN}
   
\maketitle

%________________________________________________________________

\section{Introduction}

The selection of active galactic nuclei in the hard X-ray band (up
to about 10 keV with the XMM-Newton and Chandra satellites), together with their redshift $z$
measurement after optical identification, immediately yields the
spectroscopic estimate of the photoelectrically absorbing column
density, $N_H$, intrinsic to each of them. A schematic split between objects
with $N_H$ either smaller or larger than 10$^{22}$ $H cm^{-2}$ corresponds fairly well (although not exactly one-to-one) with
the optical classification of AGN as
either type 1 (broad-line) or type 2 (narrow-line) (see, e. g., 
Perola et al. 2004). Among the objects with $N_H$ larger than 10$^{22}$ $H cm^{-2}$,
a further subdivision is necessary, those with $N_H$ smaller than $\sigma_T^{-1}=1.5\times 10^{24} cm^{-2}$, called
Compton-thin, and those with $N_H$ $>$ $\sigma_T^{-1}$, called Compton-thick.
The latter category is hardly represented in samples extracted from
images with maximum X-ray energy about 10 $keV$; therefore, the investigations
on the fraction of absorbed AGN do in fact apply only 
to the Compton-thin sources: $f_{abs} = n(10^{22} \leq N_H \leq 10^{24})/n(total)$. 

The empirical results for this fraction, reported in the literature,
have not yet reached full agreement. To first mention the two papers
(Ueda et al. 2003, La Franca et al. 2005) they
addressed the issue of $f_{abs}$ in detail simultaneously with that of the AGN
luminosity function in $L_X(2-10~keV)$, $L_X$ hereafter, and its cosmological
evolution, they agree on the existence of an anti-correlation between $f_{abs}$
and $L_X$, but not on its behaviour with $z$. While its normalization and shape stay constant,,according to
Ueda et al. (2003), according to La Franca et al. (2005, LF2005), $f_{abs}(L_X)$ increases with $z$
and becomes shallower, so that the increase is more marked at high
than at low luminosities. Akylas et al. (2006) confirm the existence
of the anti-correlation with $L_X$, but argue that the apparent increase in $f_{abs}$
with $z$ in their data sample might be due to a systematic overestimate
of $N_H$ at high redshifts. Evidence of the anti-correlation with
$L_X$ has also been reported
 by Hasinger (2004) and by Treister \& Urry (2005, see also Barger et al. 2005).
To avoid the uncertainties associated with the estimate of $N_H$,
Treister \& Urry (2006) use the optical spectroscopic classification
of the counterparts in a combination of seven X-ray-selected samples
and claim evidence (after correcting for selection bias) of an increase
with $z$ of the fraction of obscured AGN. They do not, however,explicitly address
 the issue of whether the shape of the anti-correlation might change
as well. A correlation between the fraction of broad-line 
AGN and their luminosity  is reported
by Simpson (2005), after analysing an optically selected sample ($z<$0.3)
and using the [OIII] narrow emission line to indicate the bolometric luminosity.

Although Dwelly \& Page (2006), after
a detailed study of one X-ray deep and narrow beam survey, find
no evidence of both the anti-correlation and of a redshift dependence
in $f_{abs}(L_X)$, we consider it reasonable at present to adopt the view
that the existence of the anti-correlation between $f_{abs}$
and $L_X$ is fairly well-supported.
For X-ray-selected samples in particular, we regard the results as more trustworthy
that are based on the combination of deep-narrow and
shallow-wide surveys (see LF2005 for a discussion on this point). 
On the other hand, we regard the evidence 
of an increase in $f_{abs}(L_X)$ as still controversial  with increasing $z$, even more so that of a change in its ``slope''.   
In a previous paper (Lamastra, Perola \& Matt 2006, PapI), we proposed a
simple model to explain the anti-correlation, which calls into play
interstellar gas for the Compton-thin sources, while the molecular
torus, as invoked in the standard ``unification'' model (Antonucci 1993),
remains the best explanation for the Compton-thick sources. Our proposal
is no more than an extension of the standard model, whose best empirical support
is the finding by Maiolino \& Rieke (1995) of a correlation between
absorption and disc inclination in (optically classified as) intermediate
type Seyfert galaxies, which is absent in extreme type 2 Seyferts. A finding whose
simplest explanation is that the torus and the disc are, in general, not necessarily coplanar.
The model assumes that the obscuring matter is distributed in a
rotationally supported disk, extending out to a few hundred of parsec,
as observed in nearby spiral galaxies mainly in the form of molecular
hydrogen. The gravitational pull of the central supermassive black hole
(SMBH), combined with that of the bulge (see Ferrarese \& Ford 2005
for a review of the connections between these two galactic components)
on the gaseous disc,
as a function of the distance $R$ from the centre, reduces the gaseous disc covering
factor $C$ as the SMBH mass, $M_{BH}$, increases. This effects leads to
 prediction of an anti-correlation between $f_{abs}$ and $M_{BH}$,
which is likely to be reflected into the one observed between $f_{abs}$ and $L_X$.
In PapI it was shown that the latter could be reproduced fairly well
in the ``local'' universe by adopting a ``typical'' value of 0.1,
irrespective of the mass, of the ratio between the bolometric and the Eddington
luminosity, $\lambda = L_{b}/L_{E}$. This value of $\lambda$ 
was consistent with the estimates available then  (see references in PapI).

Recently, by means of a wealth of spectroscopic data from the Sloan Digital Sky Survey
(SDSS), the empirical knowledge of the behaviour of $\lambda$ as a function
of $M_{BH}$ has had a substantial turn, thanks in particular to the work
by Netzer \& Trakhtenbrot (2007, NT2007) on broad-line AGN. They explored it
for masses from 10$^7$ to 10$^9$ $M_{\odot}$ in the $z$ interval 0.3-0.7,
and found a strong dependence on both $M_{BH}$ and $z$. It should be noted
that several earlier papers seemed to agree on a practically constant typical
value of $\lambda$ with increasing $z$, irrespective of luminosity and mass,
after selection effects had been taken into account (a good example, over the
range $z$=0.3-4 in Kollmeier et al. 2006, see also Lamastra, Matt \& Perola 2006).
In the framework of a model where, by definition, the intrinsic properties of
the central engine are the same in broad and narrow-line objects, it is natural to assume
that the behaviour found by
NT2007 holds for the narrow-line AGN as well. As a matter
of fact, in the ``local'' (around $z$=0.1) universe,
a qualitatively similar result for $\lambda(M_{BH})$ was found earlier on by 
Heckman et al. (2004, H2004) in a large sample of SDSS selected narrow-line
AGN. Despite some criticism of the reliability of the methods used by H2004
expressed by Netzer et al. (2006), we use their findings as well. 

The new light on the $M_{BH}$--$\lambda$--$z$ relationships in the range of $z$
up to 0.7 prompted us to revisit the model of PapI, with the further aim
of verifying whether the $z$-dependence of $f_{abs}$ claimed by LF2005 might
be reproduced or otherwise.

This paper is organised as follows. Section 2 is devoted to the new results
mentioned above and to their quantitative representation as is used later, Sect. 3 to the mass function of the SMBH, Sect. 4 to the molecular
gas surface density in the inner region of galaxies, both late and early-types.
Sections 5 and 6 present the model ``predictions'' of $f_{abs}$ as a function of
$L_{bol}$ and of $L_X$, and their comparison with the LF2005 results.
A discussion and a conclusion follow in Sects. 7 and 8. 

\section{New light on the $M_{BH}$--$\lambda$--$z$ relationships}

NT2007 uses a sample of almost ten thousand broad-line AGN, with $m_i \leq$ 19.1,
from the SDSS to investigate the 
distribution of $\lambda$ in different range values
of $M_{BH}$, from 10$^7$ to 10$^9 M_{\odot}$, as a function of the redshift $z$.
They estimate $M_{BH}$ using the $L_{5100}-R_{BLR}$ (the radius of the broad-line region)
relationship (as given by Kaspi et al. 2005) in combination with the FWHM($H\beta$).
The value of $\lambda$ is obtained after applying a (constant) bolometric
correction to $L_{5100}$.
The results, as summarised in their Table 1, apply to intervals of $z$
up to $z_{max}$=0.75, except for the mass range around 10$^7 M_{\odot}$
for which $z_{max}$=0.3, and down to $z_{min}$=0.05 for the mass ranges around
10$^7$ and 10$^{7.5} M_{\odot}$, with $z_{min}$ increasing up to 0.25 for
the mass range around 10$^{9} M_{\odot}$. Thus, in the $z$ interval 0.25--0.75,
the results apply to all masses from 10$^{7.5}$ to 10$^9 M_{\odot}$, while the overlap is confined to the narrow
$z$ interval 0.25--0.3 for the mass range around 10$^7 M_{\odot}$.  These results concern a feature, $\lambda_{peak}$,
defined as the maximum in the observed $\lambda$-distributions, and its dependence on 
$M_{BH}$ and on $z$. NT2007 apply reliable statistical tests to demonstrate
that the dependence found is most likely real, rather than a consequence of
the flux limit of the sample.

To avoid misunderstandings on the extent to which we rely on NT2007 in the following , it is appropriate to comment on the ``incompleteness'' of their sample.
As extensively discussed in particular by Richards et al. (2006), whose aim was to estimate 
the luminosity function of QSO as a function of the redshift, samples of such objects from the SDSS are affected
by incompleteness due to the target selection algorithm, based on the observed
photometric colors, and to the limiting magnitude. From the statements in NT2007, we infer that 
the value of $\lambda$ (see the marks in their Fig. 3, call it $\lambda_{mark}$),
below which they regard their sample incomplete, relates to the limiting magnitude $m_i \leq$ 19.1.
Figure 6 in Richards et al. (2006) shows clearly that the incompleteness degree increases very steeply for magnitudes above this limit. In the same figure it is apparent that
the incompleteness due to the color selection is of the order of 20\% (averaged   
over 15$<m_i<$19.1) and stays constant up to $z$ about 2. Going through the
way Richards et al. (2006) then used these results to obtain the proper
normalization of the luminosity function, we are led to conclude that the
color-induced selection effects do impact on the ``absolute'' values, not on the
``relative'' ones. In other words, the ``shape'' of the distributions of
$\lambda$ above $\lambda_{mark}$ should not be affected by these effects.
Since for our purpose, as will be apparent later, it is the ``shape''
of these distributions that is essential, we now describe the way we
have adopted to quantify this shape.

We introduce the distribution function $\Phi(M_{BH}, \lambda, z)$. Figure 3 
in NT2007 displays the fraction of AGN as a function of $\lambda$ for two ranges
of black hole mass (10$^{7.5}$--10$^{7.8}$; 10$^{8.5}$--10$^{8.8}$ $M_{\odot}$), each one
for three values of $z$. Except for the range M$_{BH}$=10$^{7.5}$--10$^{7.8}$ $M_{\odot}$ at $z$=0.5,  
$\lambda_{mark}$ is placed somewhat to the left of the peak in the ``observed'' distribution in the other five cases.
We thence looked for an analytical function that could represent the five ``observed'' distributions above $\lambda_{mark}$ reasonably
well. We eventually adopted a
Lorentzian profile:

\begin{equation}
\begin{array}{llllr} 
\Phi(M_{BH}, \lambda,
z)=\frac{A(M_{BH},z)}{2\pi}\frac{w}{(\lambda-\lambda_{peak})^{2}+w^{2}} &&&&\lambda > \lambda_{mark},
\label{lorentz}
\end{array}
\end{equation}
with a ``natural'' cut--off at $\lambda$=1. 

After assigning statistical error bars
corresponding to a total number of 150 objects (a relatively arbitrary choice between
the total numbers from 50 to 230 declared in NT2007 without further specifications)to each one of the five histograms in NT2007, Fig. 3,
we operated a $\chi^2$ fit to obtain the five values of $\lambda_{peak}$ and of $w$.
After noting that all five values of $w$ turned out to be equal to $\lambda_{peak}$
multiplied by approximately the same constant, we proceeded with a simultaneous fit after requiring
that this constant should be exactly the same. The result obtained is $w$ = 1.5$\lambda_{peak}$, 
with a reduced $\chi^2$=0.68. We then adopted this result to $w$ to hold for any 
mass value from 10$^{7}$ to 10$^{9}$ $M_{\odot}$, independent of $z$. It is to be noted, at last,
that our values of $\lambda_{peak}$ do not differ significantly from those,
apparently obtained by NT2007 directly from the position of the peak in their
fractional distributions, displayed in their Fig. 4. (We later return on the
normalization $A(M_{BH},z)$.)

For the $z$--dependence of $\lambda_{peak}(M_{BH})$, NT2007 find (see also
their Fig. 4) that
it can be represented (of course only in the intervals of $z$ effectively covered) 
with a power law, namely:

\begin{equation} 
\lambda_{peak} \propto z^{\gamma(M_{BH})}
\label{lambdaMbh}
\end{equation}
where the exponent $\gamma$ increases with $M_{BH}$. 
%As it can be seen in their
%Fig. 4, the power law with $\gamma$ as given in their Table 1 is followed fairly closely by
%the points representing the distribution in $z$ of $\lambda_{peak}$ 
We therefore adopt the following $z$-dependence:

\begin{equation}
\log\lambda_{peak}(M_{BH}, z) = \log a(M_{BH}) + \gamma(M_{BH})\log z.
\label{peak_M_z}
\end{equation}

\begin{figure}[h]
\begin{center}
\includegraphics[width=8 cm]{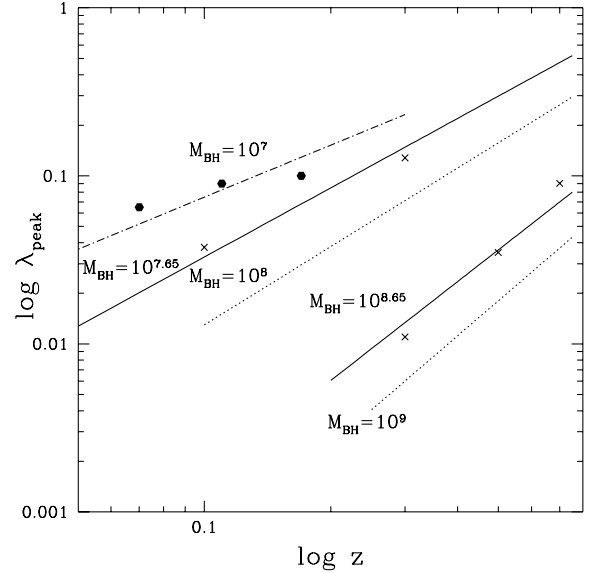}
\caption{The redshift dependence of $\lambda_{peak}$ for five values
of $M_{BH}$ in $M_{\odot}$, derived from NT2007 as described in the text.
The crosses are the values obtained by fitting the Lorentzian profile (\ref{lorentz}),
the hexagons are from Fig. 4 in NT2007}
\label{peak_z_M_Netzer}
\end{center} 
\end{figure}

For our application, it is convenient to assign an analytical form
to the function  $\gamma(M_{BH})$ and $a(M_{BH})$. For the former, it turns out that the five values of $\gamma$ in NT2007, Table 1,
are well-fitted with
\begin{equation}
\gamma(M_{BH})=-3.01+0.58\log(M_{BH}).
\label{gamma}
\end{equation}
For $a(M_{BH})$, which is not given in NT2007, we have to rely upon 
the only two values obtained with the above-mentioned fits. We choose 
to connect them with a straight line in the log-log plot of $a$ and $M_{BH}$, and obtained

\begin{equation}
a(M_{BH}) = 10^{5.5}(M_{BH}/M_{\odot})^{0.74}.
\label{a}
\end{equation}
The power-law relationship (\ref{peak_M_z}) is illustrated in Fig.~\ref{peak_z_M_Netzer} for five representative
mass values, from 10$^7$ to 10$^9$ $M_{\odot}$. Except for $M_{BH}$=10$^7$ $M_{\odot}$, the
normalization comes from (\ref{a}). The extrapolation of Eq.~(\ref{a}) to this mass places
the power law well above the three $\lambda_{peak}(z)$ points in NT2007, Fig. 4, which are
reported in our Fig.~\ref{peak_z_M_Netzer}. We therefore
decided to adopt a power law with a slope as given by Eq.~(\ref{gamma}) and a normalization 
in agreement with these points. The figure illustrates very well that the power law (\ref{peak_M_z}) 
pertaining to a given mass always stays above the one pertaining to the next higher mass. 

To use the function $\Phi$, there are three
other elements to be taken into consideration. The first would be the extrapolation of
$\Phi$ below $\lambda_{peak}$($M_{BH}, z$). This is extremely uncertain, so we avoided even trying. We exploit the fortunate circumstance that
the information available on the shape of $\Phi$ above $\lambda_{peak}$ will allow us
(see Sects. 5 and 6) to predict the quantity $f_{abs}$ for  L$_{bol}$ $\geq$ 10$^{45} erg s^{-1}$
(corresponding to $L_x$ $\simeq$ 10$^{43.5} erg s^{-1}$, Marconi et al. 2004),
namely the range of $L_x$ where the increase in $f_{abs}$ with z,is most pronounced 
according to LF2005. Incidentally, this is also the
luminosity interval where the SDSS quasar luminosity function was determined
by Richards et al. (2006); see also Hopkins et al. (2007).
 
%The first is the extrapolation
%of $\Phi$ below $\lambda_{peak}$($M_{BH}, z$). This is extremely uncertain,
%but fortunately it would impact on the predictions only at relatively low
%luminosities. For L$_{bol}$ greater than 10$^{45} erg s^{-1}$
%(corresponding to $L_x$ $\simeq$ 10$^{43.5} erg s^{-1}$, Marconi et al. 2004),
%for reasons which will become clear  in Sect. 5 and Sect 6, this extrapolation
%can be ignored. {\bf This is the reason why in the next sections the fraction
%of absorbed AGN as a function on luminosity and redshift will be given
%only for L$_{bol}$ $\geq$ 10$^{45} erg s^{-1}$, which correspond to the
%luminosty interval where the SDSS quasar luminosity function is determined
%(see Richards et al. 2006, Hopkins et al. 2007).}

The second is the normalization $A(M_{BH},z)$. Since we use 
 $\Phi$ only for $\lambda \geq \lambda_{peak}$, it is convenient to introduce
the integral quantity
\begin{equation}
\int_{\lambda_{peak}}^{1}\Phi(M_{BH}, \lambda, z)d\lambda=q(M_{BH},z)         
\label{fraction}
\end{equation}
This quantity, which depends on $A(M_{BH},z)$, represents the {\it fraction}
of $SMBH$, as a function of their mass and cosmological epoch, which are
active above $\lambda_{peak}$. 

%{\bf Thus the normalization $A(M_{BH},z)$ is
%  given by:
%\begin{equation}
%A(M_{BH},z)=\frac{2\pi q}{\arctan((1-\lambda_{peak})/w)}
%\label{Amz}
%\end{equation}
%}
There are no elements in NT2007 to infer
the value of $q$. For our purpose, however, what matters is
whether this fraction is either a constant or a function of $M_{BH}$;
in other words, whether it preserves the same {\it shape} as the active $SMBH$
mass function ($q$ = constant) or not ($q$ depends on $M_{BH}$). In Sect. 5
we illustrate the consequences of both types of assumption.

%{\bf The second is the normalization A which cannot be estimated from the NT2007
%distributions because of incompletness effects. We determined A by assuming 
%that the fraction of AGN, at a given $M_{BH}$ and z, that accrete
%at $\lambda > \lambda_{peak}$, $q(M_{BH},z)$, is the same for all $M_{BH}$ and z, i.e.:
%\begin{equation}
%$\int_{\lambda_{peak}}^{1}\Phi(M_{BH}, \lambda, z)d\lambda=q(M_{BH},z)=constant         
%\label{fraction}
%\end{equation}
%and we will discuss the
%consequences of different assumptions on the quantity q (and therefore on A) in Sect. 5.\\}

The third is the mass function (MF) of the SMBH in galaxies, $F(M_{BH})$,
in particular that of the active ones. The MF is needed
to weight the fractional function $\Phi$ when the contributions
to $f_{abs}$ by different masses are combined within the same bin
of luminosity. The MF will be dealt with in Sect. 3.

We lack a similar analysis for $\lambda(M_{BH},z)$ carried out
explicitly on broad-line AGN in the low-redshift universe. Thence
we resorted to the work by H2004 on a large (23,000) sample of
narrow-line AGN drawn from the SDSS. To the extent that this other
type of AGN can be assumed to share the same intrinsic properties
of the broad-line type, we can confidently use their results to
include the local (out to $z$=0.1) universe in our work.

Unlike NT2007, H2004 estimate the mass using the $M_{BH}-\sigma^*$
(stellar velocity dispersion in the galaxy bulge) relationship
(as given by Tremaine et al. 2002). The need to measure $\sigma^*$ 
practically excluded from the
sample the brightest Seyfert galaxies and the QSO, a cautionary
remark emphasised by H2004. They derive $L_{bol}$ from $L_{OIII}$
multiplied by a constant factor (3,500, estimated variance of 0.38 dex).
This procedure is not completely reliable, because there seems to be
a correlation between the correction factor and the luminosity
(Netzer et al. 2006), but we regard it as acceptable in the first approximation  for
our purpose. 

\begin{figure}[h]
\begin{center}
\includegraphics[width=8 cm]{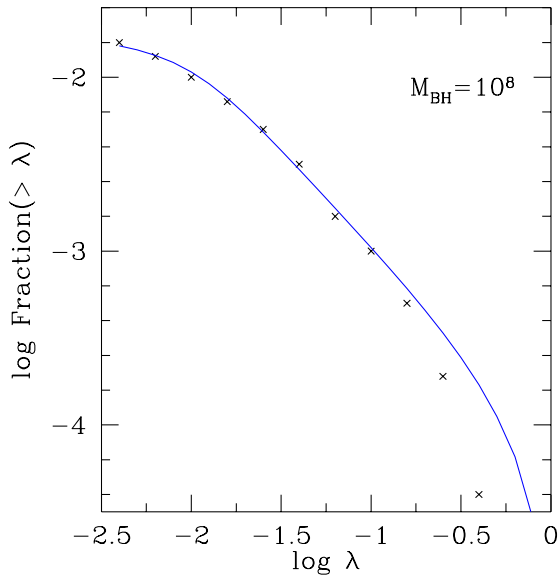}
\caption{Integral distribution in $\lambda$ of the active fraction of
SMBH, M$_{BH}$=10$^8$ $M_{\odot}$, out to $z$=0.1. The crosses are derived from H2004, Fig. 3, the
line is our Lorentzian representation.}
\label{Heck8lor}
\end{center} 
\end{figure}

The quantitative results that we  use are those summarised in H2004, Fig. 3 (left panel),
where the integral distribution in $\lambda$ of the fraction of
SMBH, which are active, is given for six values of $M_{BH}$, from
3x10$^6$ to 10$^9$ $M_{\odot}$. It can be immediately seen that,
at least qualitatively,
the trend of $\lambda$ with $M_{BH}$ is the same as found by
NT2007. This form of presentation, namely the fraction of active
objects with respect to the ``local'' $F(M_{BH})$, is particularly convenient
because it can be directly used, as we do
in Sect. 6, without 
resorting to assumptions on the quantity $q$, Eq.~(\ref{fraction}). 
It is also interesting to check whether these distributions
can be represented by the function (\ref{lorentz}). Since the curves
in H2004, Fig. 3, are devoid of error bars, we could only
adjust, as best as we could, the integral of this function to them.
The outcomes are in general quite satisfactory, and one example is 
shown in Fig.~\ref{Heck8lor} (where the curve from H2004 is represented
by a selection of points). Typically the high $\lambda$ tails stay somewhat above
the H2004 curves. One might require that the excess in the Lorentzian
tail does to some extent take care of the lack of the brightest objects
in the original sample, but we need not insist on this issue, because
the difference is of little consequence for our final results.

The parameters extracted from H2004 that we use in Sect. 6 are
summarised in Table ~\ref{tabHeckman}. The $L_{bol,0}$ in the second column
(with the corresponding $\lambda_0$ in the third) represents a completeness
limit valid for all masses between 10$^7$ and 10$^9$ $M_{\odot}$. The fourth
and fifth columns contain, repectively, $\lambda_{max}$, the maximum value
that can be read in H2004, Fig. 3, and $q_H$, the fraction of active SMBH
above $\lambda_0$ from the same figure. From the sixth column onward the quantities
given were obtained using the Lorentzian profile, as explained above.
The quantity $\lambda_{peak}$ is very low and badly determined, but the
quantity that really matters is $w$; the fraction $q_{Hl}$ in the last
column differs only marginally from $q_H$. 

It should  be noted that the fraction of active SMBH in H2004 is given with
respect to the local $F(M_{BH})$, such as the one given by Marconi et al. (2004)
and described in the next section.

%[BISOGNA INTRODURRE UN COMMENTO RELATIVO AL FATTO CHE BROAD E NARROW LINE
%AGN STANNO IN RAPPORTO CHE VARIA CON Lbol]

\begin{table*}[ht!]
\caption{The parameters used out to z=0.1, inferred from H2004}
%\begin{center}
{

\begin{tabular}{|c|c|c|c|c|c|c|c|c|}
\hline log M$_{BH}$ (M$_{\odot}$) & log L$_{bol,0}$ (erg/s) &$\lambda_{0}$ & q$_H$($\lambda > \lambda_{0}$) &
$\lambda_{max}$& $\lambda_{peak}$  & w  & q$_{Hl}$($\lambda > \lambda_{0}$) \\
\hline  7& 44.8 & 0.5 &0.32x10$^{-3}$ & 1 & 0.035 & 0.053 & 0.41x10$^{-3}$\\
\hline  7.5& 44.8 & 0.16 &1.44x10$^{-3}$ & 0.8  & 0.006 & 0.02 &1.53x10$^{-3}$\\
\hline  8& 44.8 & 0.05 &2.3x10$^{-3}$ & 0.4 & 0.005  & 0.012& 2.3x10$^{-3}$\\
\hline  8.5& 44.8 & 0.016 & 1.26x10$^{-3}$& 0.13  & 0.0005 & 0.004&1.26x10$^{-3}$\\
\hline  9& 44.8 & 0.005 & 0.17x10$^{-3}$& 0.04  & 0 & 0.0003  &0.17x10$^{-3}$\\
\hline

\end {tabular}
}
\\

\label{tabHeckman}
%\end{center}
\end {table*}

\section{The mass function of SMBH}

\begin{figure}[h]
\begin{center}
\includegraphics[width=8 cm]{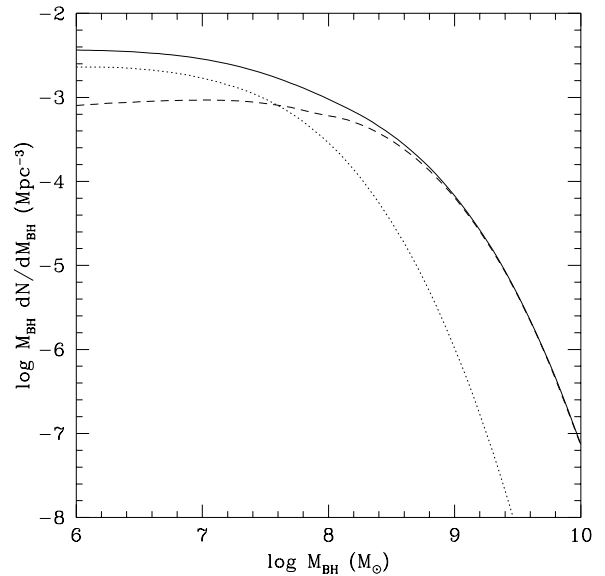}
\caption{The local SMBH mass function from Marconi et al. (2004). Dotted line: late
type galaxies; broken line: early type galaxies.}
\label{BHMF_Marconi}
\end{center} 
\end{figure}

The number density of SMBH as a function of mass,
$F(M_{BH})$, has been estimated in the local universe by several authors 
(see Marconi et al. 2004, Shankar et al. 2004 and references therein).
The one adopted here is from Marconi et al. (2004) and is shown 
in Fig.~\ref{BHMF_Marconi}, where the MF which includes all morphological types
(their Fig. 2b) is given along with the MF in late, $F_L$, and early, $F_E$, type galaxies 
separately
(A. Marconi, priv. comm.). This $F_L$ is preferred to the one estimated 
by Shankar et al. (2004) for the attention paid to the change
in the bulge to the total luminosity ratio along the Hubble sequence
of late type galaxies. The reason we keep the late separate from the early type
galaxies will be explained in Sect. 7.

For our purpose, what does matter is the shape of the MF. Insofar as
the growth of the SMBH is a consequence of accretion and is therefore
accompanied by the AGN type activity, it is most likely that the shape
of the MF for the ``active'' SMBH changes with redshift
and is different
from that of the whole population as observed in the local universe. 
In principle, the redshift dependence of the ``active'' SMBH mass function,
$F^*(M_{BH}, z)$, could be inferred from the AGN luminosity function as
a function of $z$, provided that a simple relationship holds between
$M_{BH}$ and the luminosity (for instance, $\lambda$ = constant). In practice, 
as discussed in Marconi et al. (2004) and as has become particularly evident
after NT2007, the conversion is neither straightforward
nor univocal. Hence $F^{*}(M_{BH})$ can only be reliably estimated  through
a ``direct'' evaluation of $M_{BH}$ in a properly selected sample of AGN,
with special care devoted to correcting for incompleteness selection effects.
On the basis of SDSS samples, H2004 ($z\leq 0.1$) for narrow emission line
AGN and Greene \& Ho (2007, $z\leq 0.3$, with masses derived as in NT2007) for broad emission
line AGN, both find that in the
local universe $F^{*}(M_{BH})$ is steeper, above 10$^7$ $M_{\odot}$, than $F(M_{BH})$: a difference attributed
to the so-called downsizing

In Sects. 5 and 6 we, in a first instance, adopt the ``shape'' of the
local $F(M_{BH})$; however, the impact on the results 
of other options for the downsizing is also discussed.

\section{The molecular gas surface density $\Sigma$}

The basic feature of the model in PapI is the gravitational force exerted
by the $BH$ (and by the associated stellar bulge) on a rotationally supported disc made
of molecular clouds.
This force shapes the gas distribution profile, as a function of the radial
distance $R$ from the $BH$, in such a way as to determine an anticorrelation between 
$M_{BH}$ and the ``covering factor'' determined by the disc. Quantitatively this factor 
was calculated in PapI assuming that the distance out to which the disc extends is at least
equal to 2$R_{infl}$ ($R_{infl}$ is the distance from the BH
at which the molecular gas distribution profile has a point of inflection, see Eq.~(9) in PapI), that is, about 25 pc when $M_{BH}$=10$^6$ and extending
to about 450 pc when $M_{BH}$=10$^9$ $M_{\odot}$.

\begin{figure}[h]
\begin{center}
\includegraphics[width=8 cm]{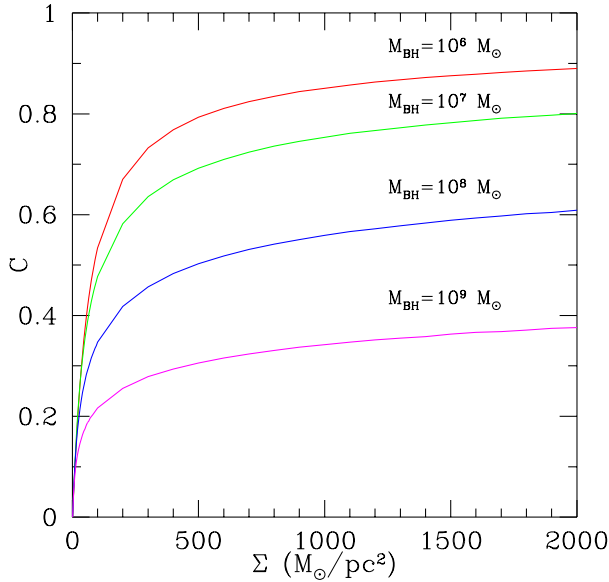}
\caption{The covering factor $C$ as a function of $\Sigma$, according to the model,
for different values of the SMBH mass. From top to bottom: 10$^6$, 10$^7$, 10$^8$, and 10$^9$ $M_{\odot}$.}
\label{Fabs_sigma}
\end{center} 
\end{figure}

Figure~\ref{Fabs_sigma} shows this ``covering factor'', $C(\Sigma, M_{BH})$, as a function of the surface density
$\Sigma$, from 0 to 2000 $M_{\odot}$ pc$^{-2}$, for four values of $M_{BH}$, from 10$^6$ to 10$^9$ $M_{\odot}$. 
After a steep growth, the curves
enter a regime where the increase is very slow. The rather sharp change in slope
takes place around a $\Sigma$ value from about 200 to about 400 $M_{\odot}$ pc$^{-2}$ 
when moving down from $M_{BH}$=10$^9$ to $M_{BH}$=10$^6$ $M_{\odot}$. Thus, in predicting $f_{abs}$,
attention must be paid to the distribution of $\Sigma$ adopted, particularly on the fraction
of objects below those values.

In PapI (to which we refer the reader for more details), we used the only
results obtained with sufficient angular resolution in a molecular line.
These are represented by a sample of 44 nearby and bright spiral galaxies
mapped at the 3 mm CO J=1--0 line, the BIMA SONG survey (Helfer et al. 2003).
At their average distance of 12 Mpc, the angular resolution of 6 arcsec
corresponds to a radial distance from the centre of about 210 pc.

\begin{figure}[h]
\begin{center}
\includegraphics[width=8 cm]{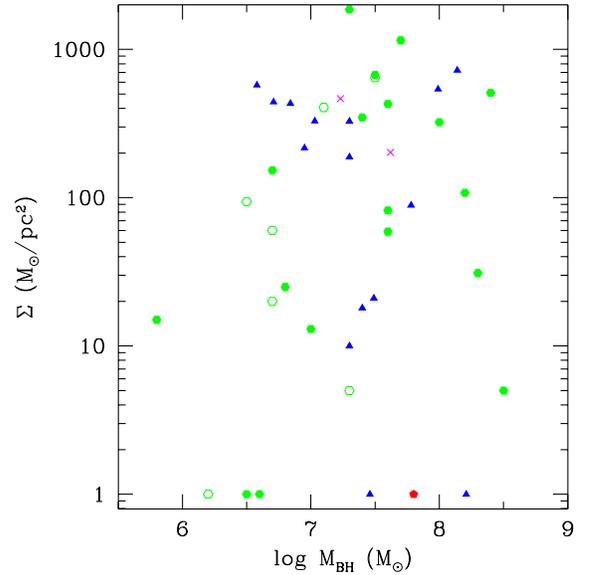}
\caption{The molecular gas central, face-on surface density $\Sigma$ of the spiral galaxies in the BIMA SONG
survey, versus the mass estimate of the SMBH hosted by these galaxies. Hexagons: $M_{BH}$
from the $M_{BH}-L_{bulge}$ relationship; open: from Dong \& De Robertis (2006); filled: 
our work. Triangles: from the $M_{BH}-\sigma^*$ relationship, Batcheldor et al. (2005), Merloni et al. (2003),
Woo \& Urry (2002), Pellegrini (2005). Crosses and pentagon: from maser and from stellar kinematics, Ho (2002).
Square: from reverberation mapping, Kaspi et al. (2000).}
\label{BIMA}
\end{center} 
\end{figure}

In Fig.~\ref{BIMA} the central face-on $\Sigma$ of these objects is plotted as a function
of the mass of the SMBH (estimated as explained in the figure
caption). For the majority of these galaxies $M_{BH}$
is comprised between 10$^{6.5}$ and 10$^{8.5}$ $M_{\odot}$. In practice this is the range
of masses for which the information described in Sect. 2 will be used to
predict $f_{abs}$ in Sects. 5 and 6. 
Given the small number of objects in this sample, we use the observed distribution
of $\Sigma$ in the 44 objects to evaluate the uncertainty on $f_{abs}$.

In PapI we stressed that the effect under study does not require any type
of correlation between the central activity and the presence of the absorbing
gas, the latter being ``in place'' as a property of the galactic interstellar medium. 
This statement can be regarded as basically correct for the late type galaxies
(not only in the local universe, as shown by the survey just mentioned, but
presumably also at cosmological distances), and now we need to consider to what extent
it might be applied to the early type galaxies.

Elliptical (E) and lenticular (S0), unlike the spiral (S) galaxies, are rare in the 
field and frequent (the bright E's in particular) in galaxy groups and clusters. Therefore, 
useful interferometric observations are only available
 for a dozen of them, optically luminous or very luminous (thus 
hosting SMBH well above 10$^7$ $M_{\odot}$), out to a maximum distance of
about 85 Mpc and with a strong CO line signal. It is convenient
to concentrate first on these results, and we  limit ourselves
to those obtained by Young (2002, 2005) with the BIMA interferometer
on seven ``true'' E (that is galaxies with an $r^{1/4}$ photometric profile).
Molecular gas is found distributed in rotating, mostly symmetric discs,
with radii ranging from 1 to 6 kpc. For each of them we report 
the peak surface brightness, translated into a surface density of H$_2$ 
gas, along with the galaxy surface area corresponding to the beam.
UGC1503 (field): 60 $M_{\odot}$ pc$^{-2}$ (2.4x2.1 kpc$^2$); NGC807 (field):  
40 $M_{\odot}$ pc$^{-2}$ (2.4x2.0 kpc$^2$); NGC3656 (merger remnant): 
720 $M_{\odot}$ pc$^{-2}$ (1.7x1.3 kpc$^2$); NGC4476 (Virgo cluster):
112 $M_{\odot}$ pc$^{-2}$ (0.73x0.49 kpc$^2$); NGC5666 (field): 
115 $M_{\odot}$ pc$^{-2}$ (1.7x1.3 kpc$^2$) (from Young 2002).
NGC83 (group): 135 $M_{\odot}$ pc$^{-2}$ (2.7x2.5 kpc$^2$);
NGC2320 (A569): 353 $M_{\odot}$ pc$^{-2}$ (2.4x2.2 kpc$^2$)
(from Young 2005). On account of the values of the area/beam,
it is evident that these selected cases correspond to the values of $\Sigma$,
which imply a substantial covering factor $C$ and which occur in 
a large fraction of the S galaxies in Fig.~\ref{BIMA}. But how frequent
the occurrence is of such a situation in E and S0 galaxies?
The answer to this question does not appear to be well 
settled yet. Furthermore, most data are from single-dish observations,
which can be properly converted only into a total H$_2$ mass.

On this issue it seems to us appropriate to first quote the
compilation by Bettoni, Galletta \& Garcia-Burillo (2003) of a wealth of data obtained
prior to the year 2003. Most useful for our purpose are this mean sample
values, subdivided per morphological class according to the mean
numerical Hubble stage index $t$ as tabulated in the Third Reference Catalogue
of Bright Galaxies (RC3 de Vaucouleurs et al. 1991), of the quantity $M(H_2)/D^2_{25}$, where $D_{25}$
is the linear size corresponding to the 25 $mag/arcsec^2$ isophote.
After assembling their classes into three groups, G1 ($t$ from -5
to -3, E), G2 ($t$ from -2 to 0, S0), and G3 ($t$ from 1 to 6, Sa to Sc),
from Table 5 in Bettoni, Galletta \& Garcia-Burillo (2003), we obtain the following 
proportions for the quantity $M(H_2)/D^2_{25}$, G1:G2:G3 = 0.06:0.29:1.
Thus, with the cautionary remark that the size of the H$_2$ gas structure 
in the three groups may not scale similarly with $D_{25}$,
the immediate conclusion one can draw is that in the ellipticals
the surface density of molecular gas is, on average, more than
one order of magnitude less than in spirals, while
it is a factor of about three less in the lenticulars. This conclusion we believe is
stronger than the one that can be drawn from the detection rate,
which is very sensitive to the sample selection, as discussed
in Combes, Young \& Bureau (2007); see also Welch \& Sage (2003), Sage, Welch \& Young. (2007)
and references in these papers. An additional important result emerging
from these papers is that, among the early type galaxies, the ratio $M(H_2)/L$ 
decreases with increasing luminosity $L$, despite the increase in depth of the potential
well. Taken together with the results quoted from Bettoni, Galletta \& Garcia-Burillo (2003),
this finding leads us to conclude that the more luminous (and therefore
hosting the more massive SMBH) an early type galaxy, the less 
the chance of finding ``in place'' H$_2$ surface densities similar
to the ones reported above from Young (2002, 2005).

For high z objects, we would like to mention the most obvious argument used to account for
the $M_{BH}-\sigma^*$ and $M_{BH}- L$ observational relationships 
(see Ferrarese \& Ford 2005 for a review), where $\sigma^*$ and $L$ are the stellar velocity dispersion
and the luminosity, respectively, of galactic ``bulges'' or ``spheroids''.
The argument goes that the formation
of the spheroids is closely connected with the growth of the SMBH sitting
at their centers.
If the two phenomena had progressed hand-in-hand 
from the earliest to the present times, it would be legitimate to postulate
that a phase of intense activity (AGN phase) should in general be accompanied by
star formation on the spatial scale of the bulge. The latter requires the 
accumulation on this scale of gas (whatever its origin) in a dense, most likely
molecular state, wherein the star formation takes place. From an observational
point of view the validity of such a postulate rests on results obtained with
studies of vast samples of galaxies, based on SDSS
spectrography and photometry. Here we refer in particular to Kauffmann et al. (2003a),
Kauffmann et al. (2003b), H2004. Quoting from the review
by Heckman \& Kauffmann (2006): ``Strong emission--line AGN inhabit those unusual galaxies
that are both relatively massive and dense, yet have a significant young stellar
population''. The denomination ``unusual'' emphasises the contrast with the
usual make-up of galaxies with similar mass and structure, which, within the limits in
sharpness of the observational distinctions employed, contain only a population
of old stars.

A cautious theoretical elaboration on these results must inevitably contemplate
the possibility that the SMBH activity is a consequence of the transformation 
into stars of the dense gas and the dynamical implications of supernovae explosions
in particular. In this instance, it could well be that, during most of the active
phase of the SMBH, the dense gas is no longer present
in sufficient quantity on the spatial scale required by our model. Furthermore,
it is doubtful that  the structure of
the molecular gas were dominated, in such a phase, even if present, by a rotationally supported configuration out
to several 100 pc. Only direct observations in molecular lines, with the
adequate angular resolution, will answer these questions.

To reach simple and sharp conclusions, we therefore consider two instances. 
In Sects. 5 and 6, in order to emphasise $\it only$
the consequences of what was described in Sect. 2, we shall posit that, for all AGN
no matter the morphological type of the host galaxies, the model applies equally well
and a single representative distribution of $\Sigma$ will be used.
In Sect. 7, we posit that the above applies only to
the late type galaxies, while in the active early type galaxies the
``covering factor'' is basically unknown. The latter instance requires
that $F_L(M_{BH})$ be kept separate from  $F_E(M_{BH})$.  
%As already stated in Sect. 3, what matters for our purpose is their
%form and relative normalization. If these two properties for the active
%$BH$ were those
%of the present day $F(M_{BH})$, the procedure would be straightforward.
%Matters might be somewhat complicated by the downsizing effect described in Sect. 3. 

\section{Predicting $f_{abs}$ as a function of luminosity for $M_{BH}$ $>$ 10$^7$ $M_{\odot}$, at $z$=0.35}

The fraction of absorbed AGN,  $f_{abs}$, in a given luminosity bin
(L$_{b1}$ $<$ L$_{b}$ $<$ L$_{b2}$)  can be computed using the equation

%\begin{equation}\label{formulafabs}
% f_{abs}(L_{b1}<L_{b}<L_{b2})=\frac{\sum_{j}\,\sum_{i}C(\Sigma_{j},M_{BH,i}) F(M_{BH,i}) \Delta M_{BH,i} q(M_{BH,i})\int_{\lambda_{1}}^{\lambda_{2}}\Phi(M_{BH,i},\lambda,z)d\lambda} {\sum_{i}F(M_{BH,i}) \Delta M_{BH,i}q(M_{BH,i}) \int_{\lambda_{1}}^{\lambda_{2}}\Phi(M_{BH,i},\lambda,z)d\lambda},
%\label{formulafabs}
%\end{equation} 

\begin{eqnarray}\label{formulafabs}
%\begin{equation}\label{formulafabs}
f_{abs}\!=\!\frac{\sum_{j}\!\sum_{i}\!C(\Sigma_{j},\!M_{BH,i})\!f(\Sigma_{j}) \Delta \Sigma_{j} F(M_{BH,i})\Delta M_{BH,i}\! \int_{\lambda_{1}}^{\lambda_{2}}\!\!\Phi(M_{BH,i},\!\lambda,\!z)d\lambda} {\sum_{i}F(M_{BH,i}) \Delta M_{BH,i} \int_{\lambda_{1}}^{\lambda_{2}}\!\Phi(M_{BH,i},\lambda,z)d\lambda}
\label{formulafabs}
%\end{equation} 
\end{eqnarray}
where $\lambda_{1}=\frac{L_{b1}}{L_{Edd}(M_{BH,i})}$ ($\lambda_{1} \geq \lambda_{peak}$) and $\lambda_{2}=\frac{L_{b2}}{L_{Edd}(M_{BH,i})}$
($\lambda_{2} \leq$ 1), $f(\Sigma_{j})$ is the fractional distribution in the quantity $\Sigma$,
introduced to take properly into account the strong dependence of $C(\Sigma)$ on
low values of the argument. In our application we used mass bins 0.3 dex wide, and the
results are presented in bins of $L_b$ that are 0.5 dex wide.A normalization of the function $\Phi$, as explained in Sect. 2, will depend on
the assumption we make on the quantity $q(M_{BH})$ given by Eq.(\ref{fraction}).
  
To predict  $f_{abs}(L_b)$ with Eq. (\ref{formulafabs}), we only consider the
active SMBH with $\lambda > \lambda_{peak}(M_{BH})$. When the relationshps (3) and (5)
are combined, it turns out that, at one particular value of the redshift, $z$=0.35,
$\lambda_{peak}(M_{BH})\propto M_{BH}^{-1}$ very closely. Therefore, all masses
in the range 10$^{7.5}$-- 10$^9$ $M_{\odot}$, to which at $z$=0.35 both (3) and (5)
apply, contribute above the same value of $L_b$, equal to 10$^{45} erg s^{-1}$.
In other words, the active $BH$ in this range of mass with $\lambda < \lambda_{peak}$
contributes only below 10$^{45} erg s^{-1}$. Notably, this is the $L_{edd}$ of
$M_{BH}$ = 10$^7$ $M_{\odot}$. Thus, to obtain a ``complete'' estimate of $f_{abs}$
in the interval of log$L_b$ = 45--45.5, only the contribution of the masses in the
range 10$^7$--10$^{7.5}$ $M_{\odot}$ needs to be added, which we did by resorting to a small
extrapolation of the curve labelled 10$^7$ $M_{\odot}$ in Fig. 1 beyond its formal limit at $z$=0.3. This convenient
situation will be exploited to make some clear point, and to predict $f_{abs}$ at a given
$z$ to be used in Sect. 6 for comparison with the predictions at higher and lower values of the redshift.

\begin{figure}[h]
\begin{center}
\includegraphics[width=8 cm]{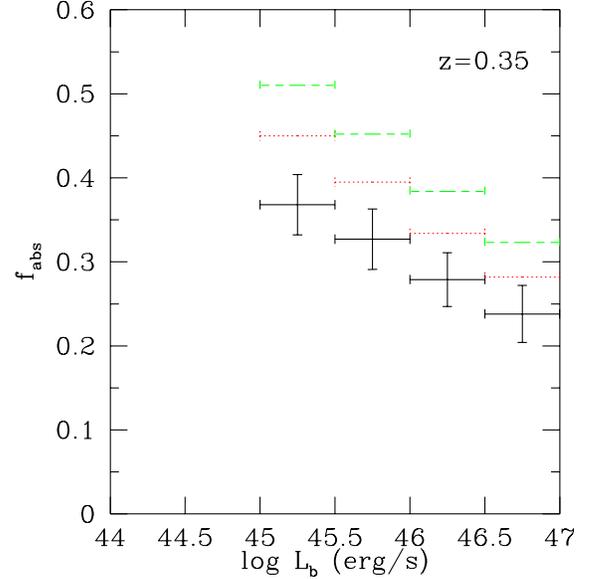}
\caption{The predicted $f_{abs}$ at $z$=0.35, obtained using the local SMBH MF, $q$=const, with three
different choices for $\Sigma$. Full line histogram: the observed distribution of $\Sigma$ described in the text,
the bars represent the systematic uncertainty due to the limited number of objects in this distribution.
Dotted and broken line histograms: $\Sigma$=150 and 270 $M_{\odot}$ pc$^{-2}$ respectively.}
\label{Abs_Lbol_sigma}
\end{center} 
\end{figure}

As anticipated in Sects. 3 and  4, we first adopt the ``shape'' of the local $F(M_{BH})$,
Fig. ~\ref{BHMF_Marconi}, as representative of the ensamble of active SMBH in both late
and early type galaxies, and assume that the ``model'' applies equally well to the
two types of hosts with the same distribution of $\Sigma$. Concerning the
quantity $q$ (Eq. ~(\ref{fraction})),
we start with the assumption that it is  equal to a constant, namely independent of $M_{BH}$.
It is perhaps worth emphasising that the value of this constant is irrelevant for obtaining the
fraction $f_{abs}$, while its precise knowledge would be necessary if we wanted to predict the number, for 
instance per unit volume, of the two types of AGN.  
The result is given 
as a histogram in Fig. ~\ref{Abs_Lbol_sigma}.
The first aspect deserving a comment is the sensitivity of the result to the choice concerning
$\Sigma$. The values of  $f_{abs}(L_b)$ in the individual bins of the histogram have been 
computed using the ``observed'', local, distribution of the 44 values of $\Sigma$ referred to
in Sect. 4, and the bars represent the 68\% confidence intervals calculated on the basis of the
limited number of objects in that distribution. For the sake of neatness, in the following 
similar histograms, the bars will no longer be reproduced. For comparison, in the same figure $f_{abs}$
is also given for two choices of a ``single'' representative value of $\Sigma$, namely the median
(150 $M_{\odot}$ pc$^{-2}$) and the mean (270 $M_{\odot}$ pc$^{-2}$) of the same distribution.
In both cases, especially in the second, the prediction is significantly higher in all bins.
This follows from the fact that (see Fig.~\ref{Fabs_sigma}), when a substantial fraction of objects
span the interval from 0 to about 300 $M_{\odot}$ pc$^{-2}$, the covering factor is extremely sensitive to 
the way the values of $\Sigma$ are distributed. In this respect, the prediction of $f_{abs}$
obtained adopting $\Sigma$=270 $M_{\odot}$ pc$^{-2}$ (as it can be judged from Fig.~\ref{Fabs_sigma})
is rather close to the ceiling attainable by pushing the distribution toward much higher values of $\Sigma$ 
than adopted here: a point on which we return in Sect. 7.

  It is important to note that, without the ``physical'' limit of unity for
  $\lambda$, the result would be different. For instance, if no
  cut-off at $\lambda$=1 in Eq. (\ref{lorentz}) is imposed, the computation of
  $f_{abs}$ results is approximately a constant value, irrespective of
  luminosity. This occurs because a single mass range,
  10$^7$-10$^{7.3}$ $M_{\odot}$, turns out to dominate at all luminosities, and the value of $f_{abs}$ to be close
  to the one pertinent to these mass values. Moreover, in this instance,  we have no information to adopt from
  NT2007 in order to estimate the
  contribution of the masses below 10$^{7}$ $M_{\odot}$, a very serious cause
  of uncertainties, due to the relatively large number of SMBH in the mass
  range 10$^6$-10$^{7}$ $M_{\odot}$. Because the relationship between $\lambda_{peak}$ and $M_{BH}$ remains
  qualitatively the same at the redshift values dealt with in the following
  section, the same type of result, with the same type of uncertainties,
  is also obtained at the other redshifts.\\

\begin{figure}[h]
\begin{center}
\includegraphics[width=6 cm,angle=270]{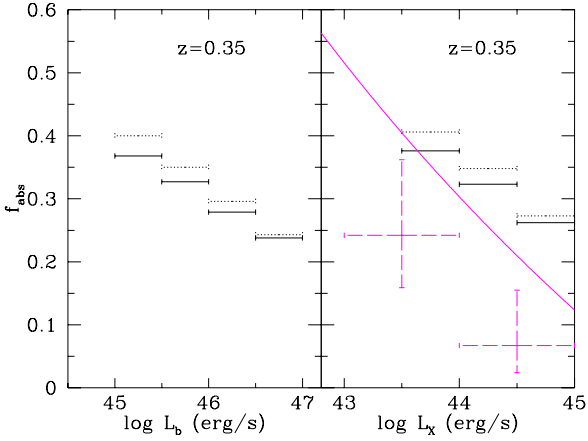}
\caption{Left panel: full line histogram, same as in Fig.~\ref{Abs_Lbol_sigma}; dotted line histogram, 
$q \propto M_{BH}^{-1}$. Right panel: histograms as in left panel as a function of $L_X$; solid
line from the global fit by LF2005; the data points, with Poissonian uncertainties, include the 
objects with $z$=0.2-0.5 in the sample used by LF2005.}
\label{Abs_z035}
\end{center} 
\end{figure}

To illustrate the consequence of taking into account the downsizing effect
mentioned in Sect. 3, we only need to change the leading assumption on $q$, namely
adopt a proper dependence of $q$ on $M_{BH}$. The shape of the broad-line AGN mass function in the local
universe, as obtained by Greene \& Ho (2007) above 10$^7$ $M_{\odot}$, suggests a dependence
of the type $q \propto (M_{BH}/10^{7})^{-0.5}$. To show that the knowledge of
the exact difference between $F^*$ and $F$ is not very important for this range of mass, we assumed 
an even steeper correction as an exercise, namely 
$q \propto M_{BH}^{-1}$. The result is shown in Fig.~\ref{Abs_z035}. 
The anticorrelation becomes somewhat steeper, but the difference overall is not particularly marked.
This is so because, even though the absolute number of objects as a function of $L_{b}$
has changed  radically, the fraction of absorbed sources depends solely
on the mix of values of $M_{BH}$ contributing to each bin of luminosity. In particular,
for instance, in the highest lumonosity bin the change is almost imperceptible
because, despite the greatest change (decrease) in the absolute number of objects,
the interval of $M_{BH}$ values contributing to it have remained almost unchanged.

To proceed to a comparison with the results obtained in the hard X-rays,
we need to turn from $L_{b}$ to $L_{x}$. The bolometric correction used comes from
Marconi et al. (2004) and is not linear. We therefore recalculated
Eq.~(\ref{formulafabs}) after converting $L_{b}$ into $L_{x}$. In the right panel of Fig.~\ref{Abs_z035}
the expectation for the two options on $q$ is compared with the results obtained by LF2005
around $z$ = 0.35. The two data points, with Poissonian error bars, represent the
objects in their samples within the interval $z$ = 0.2-0.5, while the solid line
represents the outcome at $z$=0.35 of their ``global'' fit of the intrinsic
distribution after correcting for selection effects. It
must be noted that the ``global'' fit given by LF2005 (dotted lines in their
figure 11) also includes Compton-thick AGN with 10$^{24}<N_{H}<10^{25}$, so we recalculated
their best-fit relation to take only the contribution of
the  Compton-thin AGN into account.\\
The difference between the data points and the line illustrates the
relevance of the various selection effects that LF2005 take into account
As it can be seen, our model predicts a decrease in the absorbed AGN fraction
with increasing X-ray luminosity, but the slope of the predicted
anticorrelation is not as steep as in LF2005.

\section{Redshift dependence in $f_{abs}$}

\begin{figure}[h]
\begin{center}
\includegraphics[width=6 cm,angle=270]{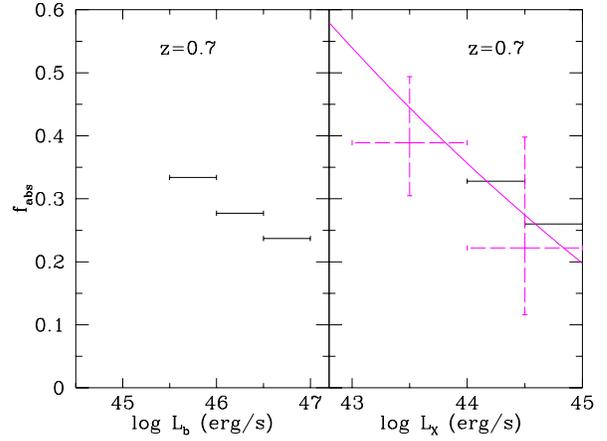}
\caption{Same as Fig.~\ref{Abs_z035}, but only with $q$=const, at $z$=0.7. The data points include the 
objects with $z$=0.55-0.85 in the sample used by LF2005.}
\label{Abs_z07}
\end{center} 
\end{figure}

We next move up to $z$=0.7, the highest redshift where the NT2007 findings do apply. The
results on $f_{abs}$ (Eq.~\ref{formulafabs}), keeping $q$ = constant, are shown in the left panel of
Fig.~\ref{Abs_z07}, as a function of $L_{b}$, for $L_{b}\geq$ 10$^{45.4} erg s^{-1}$,
where the completeness is ensured, and in the right panel of Fig.~\ref{Abs_z07}, as a function of $L_{x}$.
Notably the values of $f_{abs}$ are almost identical to those calculated at $z$=0.35 
for the same assumption on $q$. Despite the significant 
differences in $\lambda_{peak}(M_{BH})$ at the two redshifts, the parameters into play
conjure to give essentially the same quantitative outcome: not surprisingly in the
highest bin of $L_x$, which remains dominated by the highest masses. 
Even if, at this particular redshift, the global best fit at z=0.7 from LF2005
(solid line) appears to be in reasonably good agreement with our prediction, the fundamental point is that the z-dependence
of the $\lambda(M_{BH})$ distribution does not in itself introduce any
increasing trend in $f_{abs}$, at least between z=0.35 and z=0.7. 
This conclusion is strengthened by the fact that a possible dependence of $q$ on $M_{BH}$,
reflecting changes in the shape of $F^*(M_{BH})$ above 10$^7$ $M_{\odot}$, is likely to be less
steep at $z$=0.7 than assumed at $z$=0.35, for demonstrative purpose, in the previous section,
and would therefore make no significant difference for the conclusion.\\

\begin{figure}[h]
\begin{center}
\includegraphics[width=6 cm,angle=270]{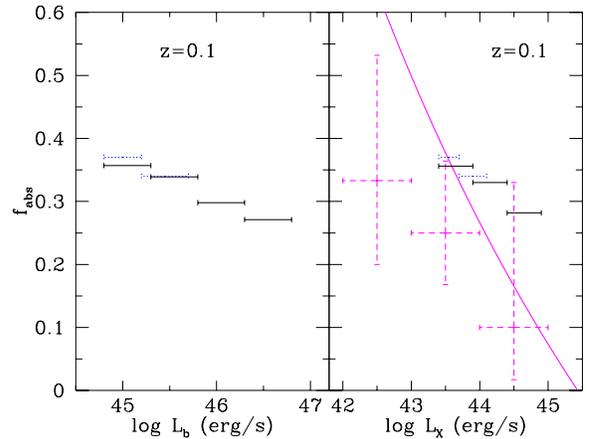}
\caption{Same as Fig.~\ref{Abs_z035} at $z$=0.1, obtained using the H2004 results, see text.
The dotted line histogram refer to the H2004 plots, the full line one to their representation
with the function given by (\ref{lorentz}).The data points include the 
objects with $z$$\leq$0.2 in the sample used by LF2005. }
\label{Heckmanz01}
\end{center} 
\end{figure}

We now move to the local Universe. We used the information extracted from H2004,
as described in Sect. 2 and summarised in Table ~\ref{tabHeckman}. Figure ~\ref{Heckmanz01},
left panel, shows the results on $f_{abs}(L_{b})$ obtained in two ways, one directly
from the curves in Fig. 3 of H2004,  the other from their Lorentzian representation.
The latter allowed us to go beyond log$L_{b}$=46, although one should be
aware that the absolute number of very luminous objects is in any
case very small in the local universe. Thus for our purpose, we consider uniportant that the H2004 sample lacks, by construction, the most
luminous objects. In the interval log$L_{b}$=44.8-46, the $f_{abs}$ expected from
the H2004 curves or from their Lorentzian representation are practically indistinguishable.

After correcting $L_{b}$ to $L_X$, the expected $f_{abs}$ is diplayed in
Fig.~\ref{Heckmanz01}, right panel. Here again the data points ($z$$\leq$0.2) 
and the best fit result
from LF2005 are also given. 
The comparison shows that the predicted slope of the anticorrelation is
definitely flatter than obtained by LF2005 at z=0.1.\\

\begin{figure}[h]
\begin{center}
\includegraphics[width=6 cm,angle=270]{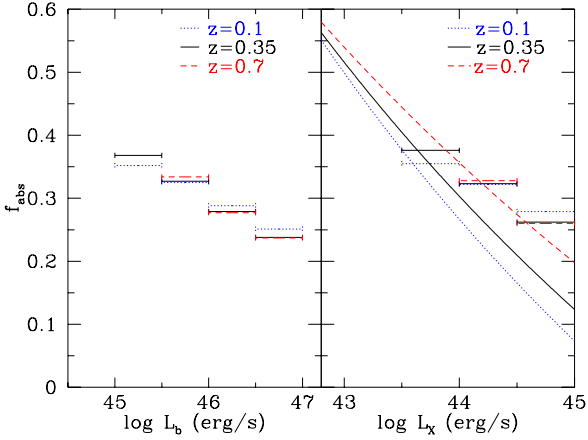}
\caption{The predicted $f_{abs}$ at $z$=0.1 and, with $q$=const, at $z$=0.35, 0.7. In the right panel the dotted, solid, and dashed lines are from the global fit by LF2005 at
z=0.1, 0.35, and z=0.7 respectively.}
\label{absoverall}
\end{center} 
\end{figure}

To summarise these results,  the $f_{abs}$ predicted as a function of $L_{b}$
and $L_X$, 
for the three values of $z$ covered (with $q$=const for z=0.35 and 0.7) are collected in Fig.~\ref{absoverall}. 
The simple conclusion that can
be drawn immediately is that, despite the strong dependence of $\lambda$
on $M_{BH}$ and on $z$, the normalization and slope of the anticorrelation
between $f_{abs}$ and $L_X$, in the range of luminosities where they
appear to change most quickly with $z$ according to LF2005,
is expected to change very little, if at all.

\begin{figure}[h]
\begin{center}
\includegraphics[width=6 cm,angle=270]{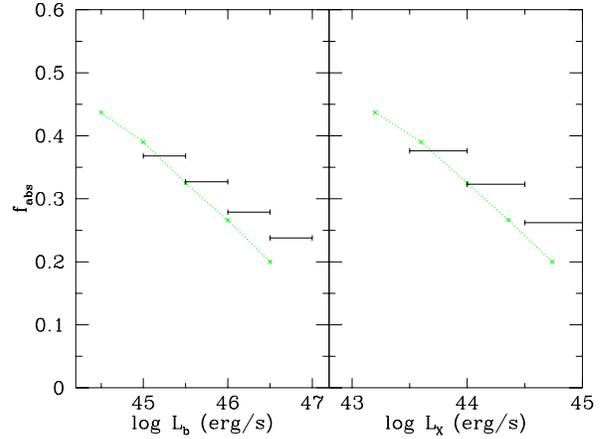}
\caption{The curve represent $f_{abs}$ at $z$ greater than 1, predicted after adopting a fixed value of
$\lambda$=0.25 irrespective of the SMBH mass. The histogram of the prediction
at $z$=0.35, $q$=const, is shown for comparison.}
\label{abszlarge}
\end{center} 
\end{figure}

We conclude this section by tentatively pushing  our predictions beyond $z$=1.
In this respect we note that, up to $z$=0.7 (see Sect. 2) the $\lambda-M_{BH}$ relationships
for the different values of $M_{BH}$ do converge. Thus we see no contradiction
with results, in particular those obtained by Kollmeier et al. (2006) and mentioned in Sect. 1,
going well beyond $z$=1, namely that out to $z$ about 4 the mean value of $\lambda$ seems to stabilize
at a value, $<$$\lambda$$>$=0.25 independent of both $M_{BH}$ and $z$, and that its distribution is
approximately lognormal. A recent investigation by Netzer et al. (2007) does essentially
confirm those results for what concerns the mean value of $\lambda$ in the redshift bin 2.3--3.4 and
the mass range 10$^{8.8}$-10$^{10.7}$ $M_{\odot}$, while the distribution is somewhat
broader than found by Kollmeier et al. (2006). A simplified approach, which ignores
the distibution and associates the same value of $\lambda$, equal to 0.25, to any
value of $M_{BH}$ $\geq$ 10$^7$ $M_{\odot}$, leads, when adopting the same $\Sigma$
distribution used previously, to the prediction of $f_{abs}$ given in Fig.~\ref{abszlarge}, which,
by construction, is independent of the SMBH mass function, and hence of $z$.
The only, albeit small, difference relative to the prediction at $z$=0.35, shown for comparison in the same figure,
is that the expected $f_{abs}$ becomes somewhat steeper, contrary to the LF2005 results.
 
\section{Discussion}

Despite the uncertainty on the actual form of $F^*(M_{BH},z)$ compared to that of the 
local $F(M_{BH})$, we have demonstrated that the predicted behaviour of $f_{abs}$ at
high luminosities does not reproduce the increase with $z$
found by LF2005, as long as the distribution of $\Sigma$ is kept fixed and indifferent
to the morphological type of the host galaxy. This result can be understood easily, 
because the trend in $\lambda(M_{BH},z)$, in particular the way this quantity for
different masses converge upwards with increasing $z$, implies that the higher the
redshift, the higher the relative contribution of the more massive SMBH (with the
lower $C$) at high luminosities. Another way to appreciate this point is to note that, 
being $L_E(10^8 M_{\odot})$=10$^{46}$ $erg s^{-1}$ ($L_{X}=10^{44.5}$ $erg s^{-1}$), beyond this 
luminosity only SMBH with higher masses contribute whose $C$ stays in the range about 0.3;
therefore, an increase in $f_{abs}$ above this range cannot be predicted.

We are therefore left to operate on the other parameter of the model, namely
$\Sigma$. On this quantity the empirical information we would need is practically
missing, and only ad hoc assumptions can be made. It should be noted immediately that,
as shown in Fig.~\ref{Abs_Lbol_sigma}, the increase in a ``typical'' value of $\Sigma$ shifts $f_{abs}$
upwards, without markedly altering its slope, in fact just rendering it somewhat steeper.
Furthermore, in Sect. 5 we emphasised that the uppermost values in Fig.~\ref{Abs_Lbol_sigma}
 practically represent a ceiling, as a consequence of the almost flat progression
of the $C$ with incresing $\Sigma$ beyond about 300 $M_{\odot} pc^{-2}$, 
irrespective of $M_{BH}$ (see Fig.~\ref{Fabs_sigma}). Thus, in the context of our model, there are limits
on the values that $f_{abs}$ can achieve, no matter the assumptions we are willing
to make.

\begin{figure}[h]
\begin{center}
\includegraphics[width=6 cm,angle=270]{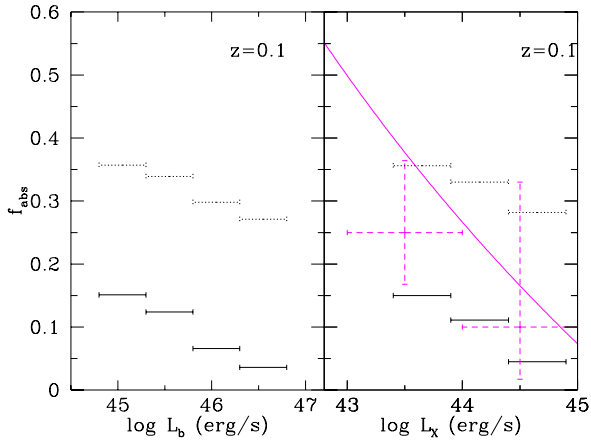}
\caption{Same as Fig.~\ref{Heckmanz01}, with the addition of the $f_{abs}$ (full line
histogram) predicted assuming $\Sigma$=0 in early type galaxies.}
\label{Abs_late_early}
\end{center} 
\end{figure}

As an exercise, let us assume that, in the local universe, the $\Sigma$ distribution
adopted in the previous two sections applies only to the late type galaxies,
the early types being comparatively, on average, much poorer in molecular gas (Sect. 4).
For simplicity, we assume that
$\Sigma$ is practically zero in the early type galaxies. Because the SMBH above 10$^{7.5-8}$ $M_{\odot}$ are
mainly hosted by this type of galaxies, above $L_X$=10$^{43.5}$ this assumption implies a major drop in $f_{abs}$,
as illustrated in Fig.~\ref{Abs_late_early} at $z$=0.1, and the discrepancy with respect to
the best fit by LF2005 is much stronger than in Fig.~\ref{Heckmanz01}.
Thence, to obtain a prediction closer in the local universe, as well
as with increasing $z$, to the results in LF2005, one would need to 
assume that the presence of obscuring gas in at least the more massive early type galaxies
(with the appropriate values of $\Sigma$
and organised as the model requires) is correlated with the nuclear activity and $\Sigma$ itself
increases with $z$. One might adopt as a hint, in favour of the latter assumption, the increase with $z$ in the
``volume-averaged'' star formation rate (SFR) (see also H2004 for further considerations).
However, this is not so straightforward, because the evolution of the X-ray luminosity function of AGN
is not only in their luminosity but also in their $\it number$ density: thus, what 
occurs in the unit volume does not necessarily apply to the individual objects.
Perhaps the simplest speculation, that one can naively propose, is that the quantity
$\Sigma$ effectively present, or at least surviving during
the activity phase, increases with $\lambda$. However, as noted above, it 
would remain hard for our model to explain values of $f_{abs}$ greater
than about 0.3 at the highest luminosities.

Unfortunately, no theoretical picture of the gas spatial distribution and kinematics,
the interplay with the SMBH growth, on the one hand, and the SFR on the other, has
been developed so far to a degree of (physically consistent) detail necessary 
to sensitively adjust the prediction of $f_{abs}$. On the observational side, we lack
sufficient data to proceed empirically, as we have done with $\lambda$.

We finally need to point out the non-negligible difference between the optical
and the X-ray selection of AGN. Following Rigby et al. (2006) (see also Fiore et al. 2003, 
Cocchia et al. 2007 and references in all three papers), we stress that a 
substantial fraction of AGN selected through their X-ray emission and
classified as such on the basis of their $L_{X}$ would be missed through
an optical spectroscopic classification (this fraction increases from about
15\% locally to about 50\% at $z$ around 1). These objects could perhaps
reduce the difference between our predictions and the $f_{abs}(z)$ found
by LF2005. Such a possibility could be tested only after a properly complete
investigation of the $\lambda$ of these objects is carried out.

\section{Conclusions}

In the light of recent new results on the behaviour of the Eddington ratio $\lambda$
as a function of the SMBH mass and of the redshift, we have revisited the
predictions of the model presented in PapI. This model naturally predicts 
an anti-correlation between the fraction $f_{abs}$ of X-ray
obscured, Compton-thin AGN, and $M_{BH}$, provided there is interstellar
molecular gas with appropriate values of surface density $\Sigma$ approximately within 
 25 to 450 pc (depending on $M_{BH}$) from the active nucleus.
We used the newly found $\lambda(M_{BH},z)$ dependence ($z$ from the local
universe to 0.7) and a distribution of $\Sigma$, irrespective of the
morphological type of the host galaxies and invariant with $z$, 
to convert that prediction into one $\it versus$ $L_X$. The results
appear to reproduce, at least qualitatively,
the anti-correlation between $f_{abs}$ and $L_X$ claimed by several
authors, but not the still controversial
claim of an increase in $f_{abs}$ with $z$.
The latter increase, particularly in the form claimed by LF2005 on which we
did concentrate, might possibly be reproduced by assuming that the typical
values of $\Sigma$ do increase with the redshift and,
furthermore, that in the early type galaxies, at least in the more
massive ones, the presence of this quantity is correlated in time with
the activity of their nucleus. At the highest luminosities, however,
according to our model, $f_{abs}$ can hardly exceed 0.3.\\    
For the sake of completeness, it should be noted that the results just summarised are sensitive to the presence of a cut-off in the $\lambda(M_{BH},z)$ distribution, and they correspond to the case where the cut-off is placed at the physical limit $\lambda$=1.\\

Steps forward, that are relevant not only to the model discussed in this
paper can be made at least in two ways. The first and most obvious one 
requires  systematic investigation of the amount of molecular gas in
an appropriately selected sample of early type galaxies, both active
and inactive, out to redhifts of 0.5 at least. The line strength and profile may not tell us enough about
the spatial distribution, but much about the amount and the kinematics 
of the gas. The former property needs to wait for observations with a large
interferometric array of telescope, such as ALMA. The second step is
somewhat indirect and therefore requires interpretation. The method 
is described and applied to a handful of broad-line objects in Maiolino et al. (2007).
It allows an estimate of $C$ to be inferred for each object on the basis of 
measurements of infrared spectral features due to dust reradiation
of the ultraviolet from the active nucleus. It is worth noting that
Maiolino et al. (2007) find an anti-correlation between $C$ and either
the luminosity or the mass, thus leaving completely open at present
which of the two parameters is the one ``physically'' driving
the anti-correlation. This sort of degeneracy could be resolved by
comparing the covering factor of objects with similar $M_{BH}$
and widely different values of $L_b$, hence of $\lambda$. According to our model
a similar value of $C$ should be found in such objects.

\section*{Acknowledgements}
The authors are particularly grateful to Fabio La Franca for his help in assessing
the figures with the X-ray data, and thank several colleagues for enlightening
discussions. 
They acknowledge financial support from the ASI (grant 1/023/05/0) and MiUR (2006025203).

\end{document}